# Can we illuminate our cities and (still) see the stars?

Salvador Bará[1,*], Fabio Falchi[1,2], Raul C. Lima[3,4], Martin Pawley[5]

[1]Departamento de Física Aplicada, Universidade de Santiago de Compostela, 15782 Santiago de Compostela, Galicia
[2]Istituto di Scienza e Tecnologia dell'Inquinamento Luminoso (ISTIL), 36016 Thiene, Italy.
[3]Escola Superior de Saúde do Politécnico do Porto, 4200-072 Porto, Portugal
[4]Instituto de Astrofísica e Ciências do Espaço (IA), Universidade de Coimbra, PT3040-004, Coimbra, Portugal
[5]Agrupación Astronómica Coruñesa Ío, 15005 A Coruña, Galicia

**Abstract**

Could we enjoy starry skies in our cities again? Arguably yes. The actual number of visible stars will depend, among other factors, on the spatial density of the overall city light emissions. In this paper it is shown that reasonably dark skies could be achieved in urban settings, even at the center of large metropolitan areas, if the light emissions are kept within admissible levels and direct glare from the light sources is avoided. These results may support the adoption of science-informed, democratic public decisions on the use of light in our municipalities, with the goal of recovering the possibility of contemplating the night sky everywhere in our planet.

## 1 Introduction

A common tenet of the dark skies movement is to try to preserve the dark areas still existing in the world, avoiding their further deterioration. This position, often implemented through different figures of protection like certified Starlight Tourist Destinations [1-2], IDA's International Dark Sky Places [3], and other astro-tourism initiatives [4-8], not only intended for pristine dark sites, but also addressing urban parks and star observing spots in polluted areas, has brought extremely significant benefits to the cause of the night and should no doubt be further promoted. Notwithstanding that, a purely defensive, reactive stance seems to be nowadays insufficient for ensuring the future of the dark nights in the planet. The persistent increase of radiance and illuminated surface [9] progressively encroaches the dark areas, reducing their size and natural values [10-12]. The dark territory to defend becomes progressively smaller and fragmentary, breaking in many cases the continuity of the nocturnal ecological corridors [13].

A related ethical and political issue is whether or not we should collectively renounce to restoring the quality of the night in areas already deteriorated, particularly in the most conspicuous ones, our urban nuclei. It is often taken for granted that the night skies are irremediably lost there, due to the huge amount of light produced in our urban agglomerations. Conventional wisdom considers the present level of emissions unavoidable. There is however an ongoing trend to reassess the validity of some "old truths" regarding the required light levels in cities: modern research fails once and again to find convincing reasons to support the actual recommendations for road lighting [14], be it in the name of a purported traffic safety, fostering of compulsive consumption, or the even less proven effect of increased photons densities when it comes to avoid some behaviors, see e.g. [15-20]. Recovering the night sky in our metropolitan areas should therefore not be discarded a priori. The explicit position of this short paper is that decidedly remediating deteriorated skies should be the option by default, and that this goal should not be abandoned excepting where and when proven unfeasible.

*S. Bará, E-mail address: salva.bara@usc.gal

Assuming that a significant amount of light will be used anyway in populated areas, a natural question arises: which is the maximum level of emissions compatible with ensuring a given darkness of the urban night sky? Which are the compromises and balances? This work provides some general estimates of the order of magnitude of the maximum allowable emissions, with the aim of contributing to the open discussion of this issue. Establishing the "red-lines" [21] for the celestial nightscapes that we collectively desire for the nights of the places where we live is a necessary pre-requisite for negotiating these goals against other legitimate social wishes.

## 2 Seeing the stars

Several well-known factors limit the ability of the human visual system to detect the faintest visible stars. Some of them are directly related to the eye optics and the first steps of neural image processing that determine the minimum luminance required to detect a beam of light, as well as the luminance contrast thresholds to identify the presence of a celestial object against a lit background. The overall performance of our visual system results from the interplay of the quantum efficiency of the retinal photoreceptors, the photon noise, the pre-processing of the detected signals by the first post-receptor layers of the retina (horizontal, bipolar, amacrine, and ganglion cells) and the subsequent cortical processing [22]. The fraction of incoming radiation entering the eye and reaching the individual photoreceptors in the retina is itself dependent on the pupil size and spectral transmittance of the ocular media of the observer (both age-related), as well as on the presence or not of residual uncorrected ametropia and on the typical size of the ocular point-spread function associated to the physiological aberrations of the healthy human eye [23-24]. Last but not least, the use or not of binocular vision, and the experience of the observer may significantly influence their ability to detect the faintest stars.

For many practical applications all these factors can be summarized in a single number, namely the luminance contrast threshold required to detect the presence of a stimulus of a given angular size against a more or less bright background, under the prevailing observing conditions [25]. The concept of luminance contrast threshold is key for assessing the possibility of detecting stars. Be it defined as a Weber fraction or as a Michelson contrast, the existence of this threshold requires that, as the background luminance increases, the object luminance must also increase –depending on the luminance adaptation state of the eye, and not necessarily in a linear way– to be in the limit of detection. Any increase of the luminance of the background will then result in a limitation of our ability to detect faint stars.

The natural sky background over which we see the stars is by no means absolutely dark in the optical region of the spectrum, even in moonless nights during the astronomical night. Besides the unresolved stars of the Milky Way, it has contributions from the galactic and extragalactic diffuse light, the light scattered by the zodiacal cloud and, especially, the highly variable atmospheric airglow [26]. After entering the Earth's atmosphere part of this light is scattered by molecules and aerosols, contributing that way to increase the visual background (or foreground, perhaps a more appropriate denomination in this case). On the other hand, the direct radiance of the sky objects propagated along the line of sight is attenuated by the atmosphere through the complementary mechanisms of scattering out of the beam and absorption along the path, thus contributing to a further reduction of the contrast of the stars. Everything else being equal, the natural sky is brighter at greater elevations above sea level, and the number of visible stars shall decrease as the observers approach the sea baseline and/or look towards larger zenith angles, as shown by Cinzano and Falchi [27]. All these factors can be quantified, for a given observing site, time, and atmospheric conditions, by means of the Gaia-Hipparcos multi-band map of the natural night sky brightness [26], and its associated GAMBONS web tool [28]. Needless to say, all of these background light sources and attenuation effects are themselves an integral part of the natural sky we seek to preserve or recover, not an effect to avoid or remediate.

Artificial light sources, in turn, add scattered photons and in some cases also direct ones to the observed scene, further reducing the contrasts. It is not always fully realized that this artificial sky brightness is not generated "in the sky" or in some particular layer of the atmosphere high above our heads, but it is a highly distributed phenomenon due to the cumulative contributions of photons scattered across the air column along our line of vision, starting at the very first millimeter in front of our eyes. Depending on the location of the sources, their angular and spectral emission patterns, and the aerosol types and concentration profiles, the maximum contribution to the total amount of atmospheric scattered radiation entering our eyes will be generated at some definite altitude above ground, but all path length elements have definitely their share, from millimeter one, so to speak. Additionally, the radiance directly entering our eyes from directions in the periphery of the visual field (e.g. from unshielded streetlights) is strongly scattered in the intra-ocular media [29-30], giving rise to an additional large number of photons superimposed onto the photoreceptor receptive fields that further reduce the contrast, a phenomenon usually termed glare. Finally, an additional extra-atmospheric source of background artificial light is the diffuse radiance of the satellite and space debris cloud [31].

All of the above put together, it is possible to formulate reasonably comprehensive and accurate models for predicting the limiting magnitudes of the stars that can be seen by humans under a variety of observing conditions. Detailed formulations can be found in the classical works by Shaeffer [32-34], which are the key references in this field, and in other related publications [35-37].

A proxy commonly used in the light pollution research community for reporting the visual quality of the night sky is the total sky brightness in the Johnson $V$ band, specified in $mag_V/arcsec^2$. The use of the Johnson $V$ band [38-39] is partly motivated by historical reasons and should probably be revised in the near future. From a visual point of view the Johnson $V$ band turns out to be suboptimal: the sky luminance in the photopic [40], scotopic [41], or mesopic [42-43] human vision bands should be routinely reported instead, whereas for large-scale, worldwide light pollution instrument-based measurement and reporting campaigns the scientific photometric system defined for the standard RGB bands [44], such as those used in science-, industry-, and consumer-grade digital cameras [45], seems to be the-most efficient choice. For the purposes of this work, the quality of the urban night sky will be characterized with two metrics commonly used in this field: the total zenith sky brightness in Johnson $magV/arcsec^2$, and the photopic artificial zenith luminance, in $cd/m^2$. The latter will be natively evaluated in the CIE photopic $V(\lambda)$ band, i.e. not transformed from the Johnson $V$ radiance (for the issues associated with this transformation see [46-47]). The use of these well-known metrics is intended to facilitate a first discussion of the issue of the restoration of urban skies. More comprehensive metrics, including all-sky and near-horizon average brightnesses [48-50] should be taken into account in future works for a full formulation of the desired features of the urban nightscapes.

## 3 City emissions and artificial sky brightness

### *3.1 Model and parameters*

To answer the question that motivates this paper it is convenient to analyze a simplified situation that nevertheless captures the basic physics of the problem. There is no fundamental difficulty for computing more accurate results, tailored for any actual urban configuration one may wish, and no doubt these calculations should be performed in any particular city when it comes to adopting science-informed public decisions on lighting. However, such a level of detail does not seem to be necessary for giving an order-of-magnitude answer to the issue addressed here.

Let us then consider a plausible scenario: an observer located within small urban park of radius, say, $R_0$~200 m, embedded in the center of a wide metropolitan area of arbitrary radius $R$. To avoid short-distance scattering effects as well as the glare due to the direct radiance of the streetlights entering the eyes of the observer, let us assume that the park is reasonably free from annoying artificial light sources. This does not preclude the existence of light sources to illuminate the pedestrian paths, as long as their light –including the one reflected off the ground– is neither redirected in appreciable amounts towards the sky nor towards the observer eyes, being perhaps blocked by the tree canopy.

Let us further assume that the city light emissions are spatially homogeneous in average, that is, that the amount of artificial light emitted per unit of city surface, $\Phi_L$, measured in $lm{\cdot}m^{-2}$ (or, equivalently, in $Mlm{\cdot}km^{-2}$) is reasonably constant across the metropolitan area. Under these assumptions, and with some additional information, the zenith brightness at the observer location can be easily calculated from the spectral power density of the city lights by using the extremely simplified model described in the Appendix, eq. (A12). Different approaches can be used to compute the required point-spread functions (PSF) that describe the propagation of light from the street lamps to the observer [51-57]. For illustrating the examples below, we used the Illumina v2.0 spectral PSFs calculated by Simoneau et al (2021), available in the excellent paper [58]. These PSFs are given in [58] as a function of the distance to the source for different wavelengths, specific luminaire angular emission patterns with different fractions of light directly emitted to the upper hemisphere (ULOR), obstacle heights (h), and aerosol optical depths (AOD), keeping constant other parameters as the luminaire pole height, street form factor, spectral reflectance of the terrain, and the atmospheric conditions other than AOD, e. g., the aerosol composition, albedo and scattering phase functions. Previous versions of Illumina PSFs (v1.0) had been successfully used to compute the artificial sky brightness in other settings, see, e.g. [59-60].

The Johnson $V$ photometric system adopted here is the one described in [26], using the classical Bessell passband and a Vega-based magnitude scale in which $mag_{Vega}$=+0.03, which corresponds to a zero-point reference radiance $B_{ref} = 143.1682$ W·m$^{-2}$·sr$^{-1}$. The brightness of the natural sky in the Johnson $V$ band is an important parameter for interpreting the measured sky brightness, which is the sum of the natural and artificial components. The natural brightness of the zenith night sky adopted here is $m_N = 22.0$ magV/arcsec$^2$, a value conventionally used for this kind of studies [10]. Note that the zenith natural brightness is actually highly variable in time, since it depends on the patch of the sky located around the zenith region at each moment, as well as on the intensity of the airglow and the overall state of the atmosphere. The interested readers may want to use slightly different values for their locations, times, and atmospheric conditions, information freely available in the GAMBONS website [28]. The GAMBONS tool has also been used here to determine a reasonable value for the visual luminance of the natural sky, set to 200 mcd/m$^2$. This reference natural luminance, determined natively in the CIE photopic visual band by using the Gaia-Hipparcos map of the brightness of the natural sky [26], is consistent with the values determined in an indirect way, namely by performing appropriate conversions from the Johnson $V$ band radiance [46-47], and it should be recommended as a default value instead of the less realistic 174 mcd/m$^2$ frequently quoted in the literature.

### *3.2 Results*

Figure 1 displays the total (natural + artificial) zenith brightness of the sky in the Johnson V band, expressed in magV/arcsec$^2$ (Fig 1a) as well as the artificial component of the photopic zenith luminance given in mcd/m$^2$ =10–3 cd/m$^2$ (Fig 1b), for an urban area whose light sources are entirely composed by 3000 K CCT LED, typical obstacle height 9 m, ULOR=5% and an atmosphere with an aerosol optical depth (AOD) of 0.2, corresponding to a visibility of about 26 km. The values associated to the different lines shown in the legend (1, 2, 3, 5, 10), correspond to different levels of city light emissions spatial densities, $\Phi_L$, expressed in lm·m$^{-2}$. These values can be converted to average illuminances at street level, $E$, in lx, by using the formula $E = \eta\Phi_L/\epsilon$ where $\eta = 1 - \text{ULOR}$ is the fraction of radiant flux emitted towards the lower hemisphere, and $\epsilon$ is the fraction of the city territory actually illuminated, i.e. excluding roofs and other zones where no direct light arrives. Given the relatively low ULOR value of the sources considered in this figure, the average street illuminance is approximately equal to $\sim\Phi_L/\epsilon$ to within a 5%, such that if the outdoor spaces directly lit occupy, say, only one half of the city surface, $\epsilon = 0.5$, then the average street illuminance (in lx) compatible with attaining a given level of sky darkness (in $mag_V$/arcsec$^2$) will be twice the number indicated in the legend.

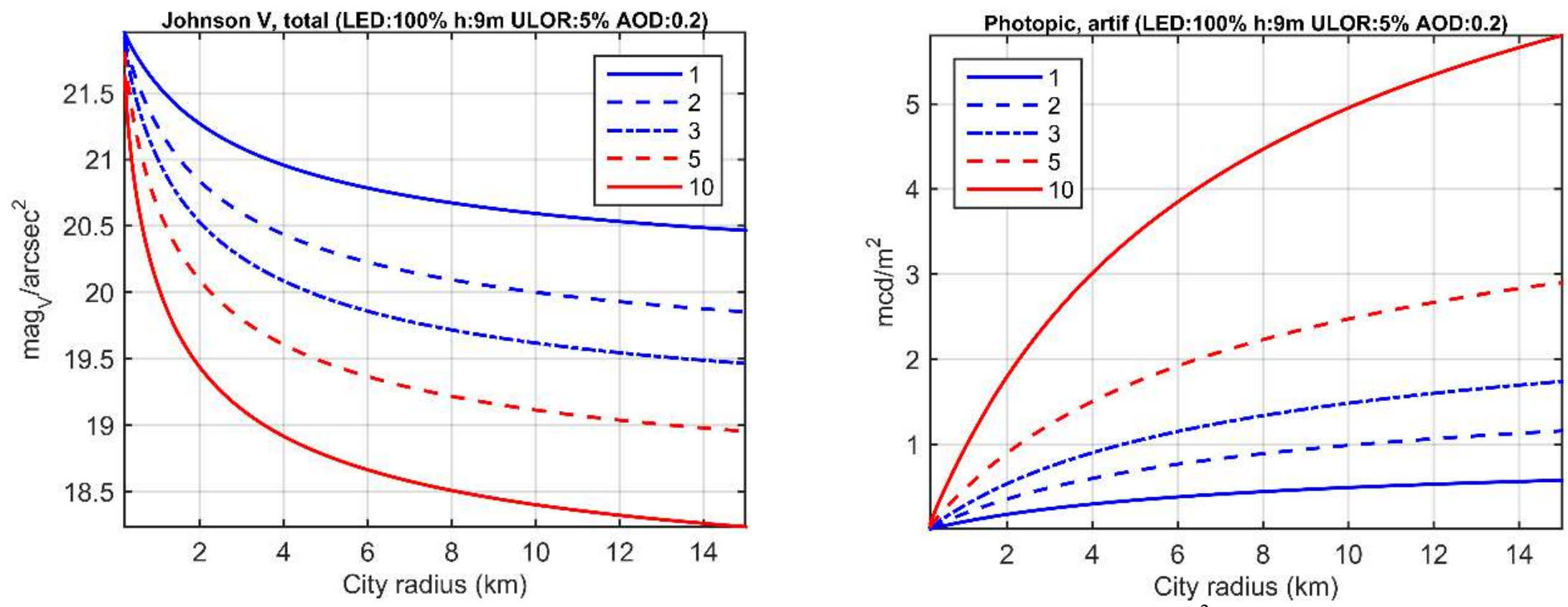

Figure 1: Brightness of the zenith sky for different amounts of emitted artificial light (legend in lm·m$^{-2}$), versus city size. Left: total brightness (including the natural component) in the Johnson V band, in magV/arcsec$^2$. Right: artificial component luminance in mcd/m$^2$. See text for details.

According to this figure and under the assumed conditions, a city of radius 4 km, with $\epsilon = 0.5$ could in principle attain zenith skies slightly darker than 20.0 magV/arcsec$^2$ if its outdoor spaces are lit with an average illuminance not surpassing 6 lx (blue dash-dotted line of emissions 3 lm·m$^{-2}$). Achieving 21.0 magV/arcsec$^2$ is definitely possible if the overall emissions do not surpass 1 lm·m$^{-2}$. Keep in mind that these results are contingent on the assumption of an isolated city, such that no other urban area adds polluting photons to the sky of the observer. As the size of the city increases, the emission densities compatible with a given darkness of the sky tend to decrease, due to the additional light contributions from the city rings of increasing area $r\,\mathrm{d}r$ located at progressively larger distances from the observer. Note, however, that the contributions from larger $r$ are exponentially attenuated by the atmosphere, so the total brightness increases at a progressively smaller rate with city size.

The actual brightness of the sky is of course highly variable along the night and throughout the seasons, depending on the evolution of the artificial light emissions and the continuously changing atmospheric conditions. Cities with different urban structures are expected to experience different levels of sky brightness, due to the particular types of angular emission patterns and spectra of their light sources, the different kind of obstacles and the specific spectral reflectance of their surfaces. To get some insights about the variability associated to some of these factors, the following figures display the total sky brightness in the Johnson V band and the artificial component of the zenith sky luminance for several combinations of them.

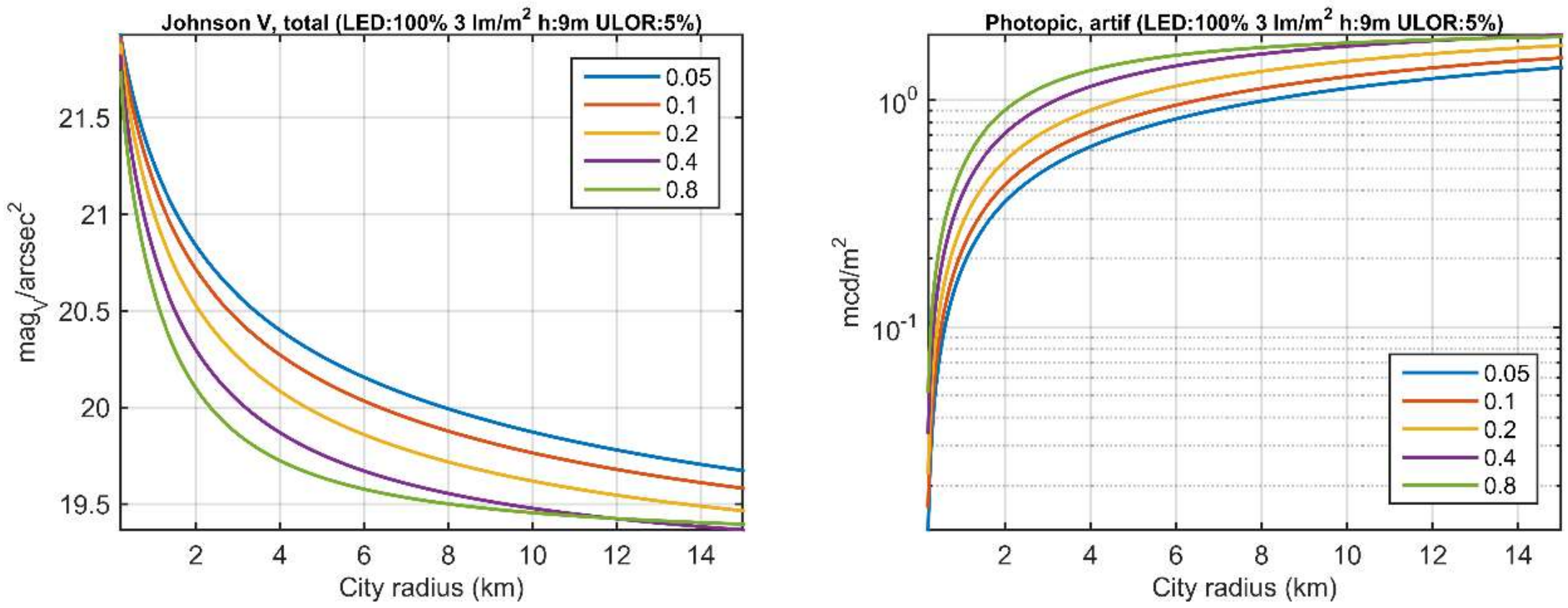

Figure 2: Brightness of the zenith sky for different AOD (legend, unitless), versus city size. Left: total brightness (including the natural component) in the Johnson V band, in magV/arcsec$^2$. Right: artificial component luminance in mcd/m$^2$. See text for details.

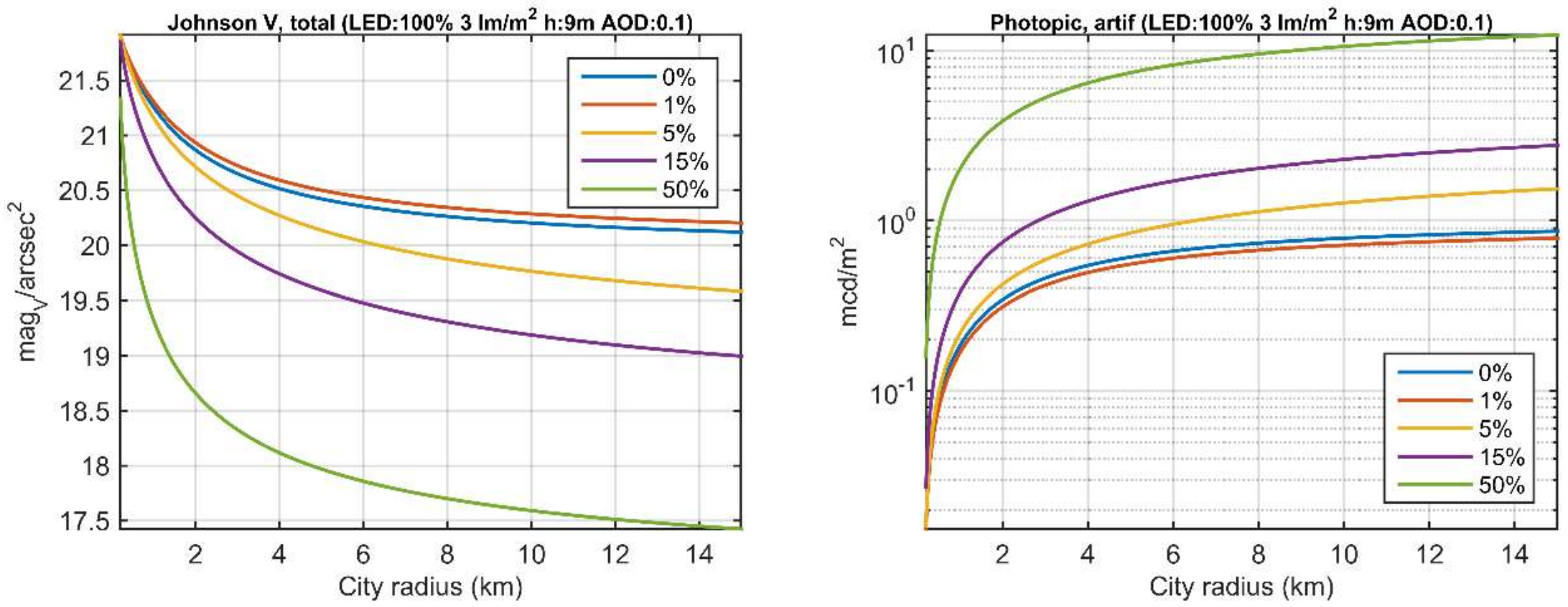

Figure 3: Brightness of the zenith sky for different ULOR (legend, %), versus city size. Left: total brightness (including the natural component) in the Johnson V band, in magV/arcsec$^2$. Right: artificial component luminance in mcd/m$^2$. See text for details.

Figure 2 shows the influence of the aerosol optical depth, ranging from 0.05 (transparent nights) to 0.8 (turbid ones). As expected from the dynamics of the light propagation in the vicinity of the sources, thicker optical atmospheres tend to give rise to brighter artificial skies, due to the larger fraction of scattered light, which is determined by the local aerosol concentration. The absolute amount of scattered light is also proportional to the amount of incident light, that gets attenuated by propagation through the atmosphere. For nearby sources, as is the case analyzed here, larger scattering efficiencies tend to compensate for larger attenuations. Note however that for very thick atmospheres the attenuation of the light from the distant city districts may result in darker skies in comparison with thinner ones (see the intersection of the curves corresponding to AOD 0.4 and 0.8 for cities of radius larger than ~12 km).

As suggested by Figure 2, reducing the aerosol content above cities reduces the amount of scattered light and hence allows to enjoy darker skies. Since an important part of the urban aerosol content is due to the air pollutants produced by human activities, mitigating air pollution in cities is not only necessary in terms of public health but also provides, as a positive side-effect, night skies where more stars can be visible to the unaided eye. This association has been thoroughly analyzed in a recent paper by Kocifaj and Barentine [61]. Their results nicely complement the ones reported by Stark et al. [62] that established an association in the reverse sense, showing that emissions of artificial light slow down the natural reduction of the concentrations of some air pollutants at nighttime by means of photo-chemical reactions that prevent their degradation.

Figure 3 displays the zenith sky brightness associated to the use of luminaires with different ULOR (fractions of flux directly emitted to the upper hemisphere, in %). The artificial sky brightness is higher for larger ULOR, as anticipated, excepting for the cases 0%-1% in which a curious inversion of this trend can be seen because the obstacles are higher than the lights, as analyzed in Simoneau et al [58].

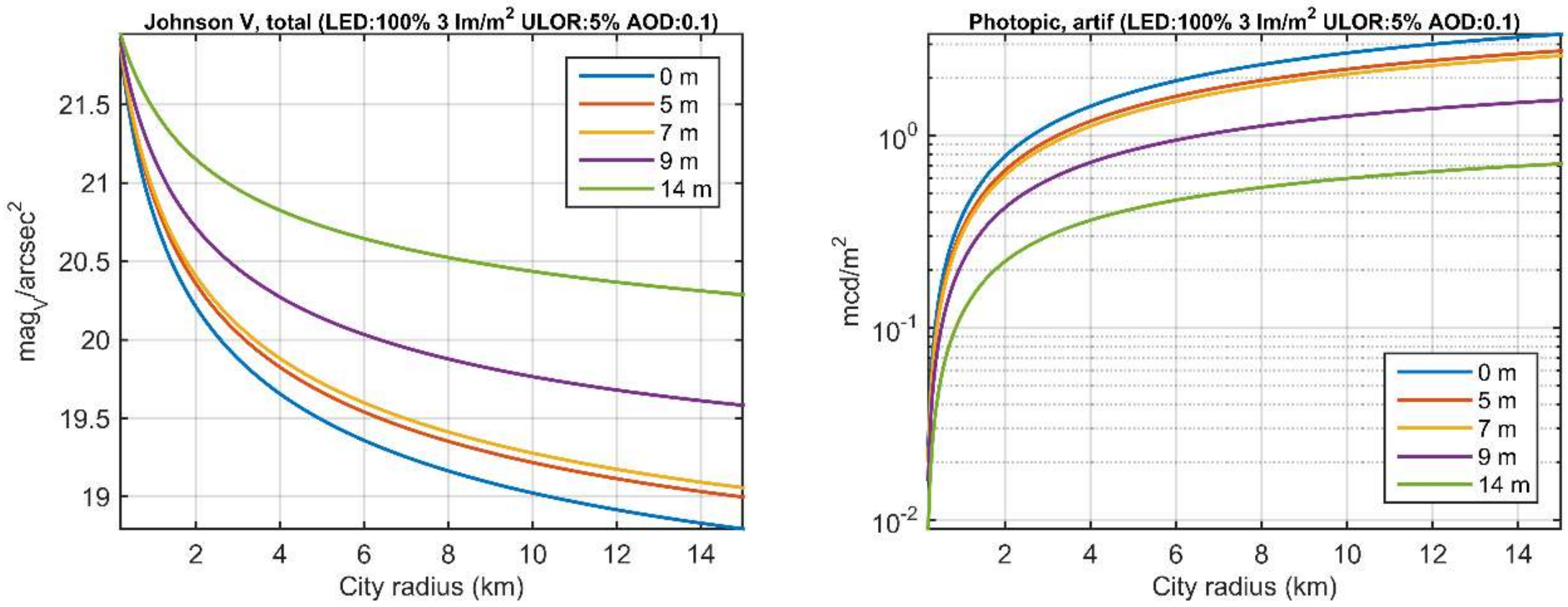

Figure 4: Brightness of the zenith sky for different obstacle heights (legend, m), versus city size. Left: total brightness (including the natural component) in the Johnson V band, in magV/arcsec$^2$. Right: artificial component luminance in mcd/m$^2$. See text for details.

The influence of the obstacle height is depicted in Figure 4. Distance between obstacles is assumed to be 20 m, obstacle filling factor 80 % and lamp height 7 m, according to Table 1 of [58]. As expected, cities with smaller height-to-width street shape factors have significatively more polluted skies for the same light emission densities than cities with high buildings in narrow streets. This is due to the fact that the light emitted at low angles above the horizon is not blocked and contributes efficiently to the brightness at the observer location. It is worth recalling that the spatial structure of the urban nuclei is highly variable across the world, depending on multiple factors that include land availability, land property rules, building regulations, and the historical evolution of the building markets. It is this structure, together with the angular emission pattern of the luminaires and the bi-directional reflectance distribution function of the pavements and façades, what ultimately detemines the angular emission pattern of the city as a whole.

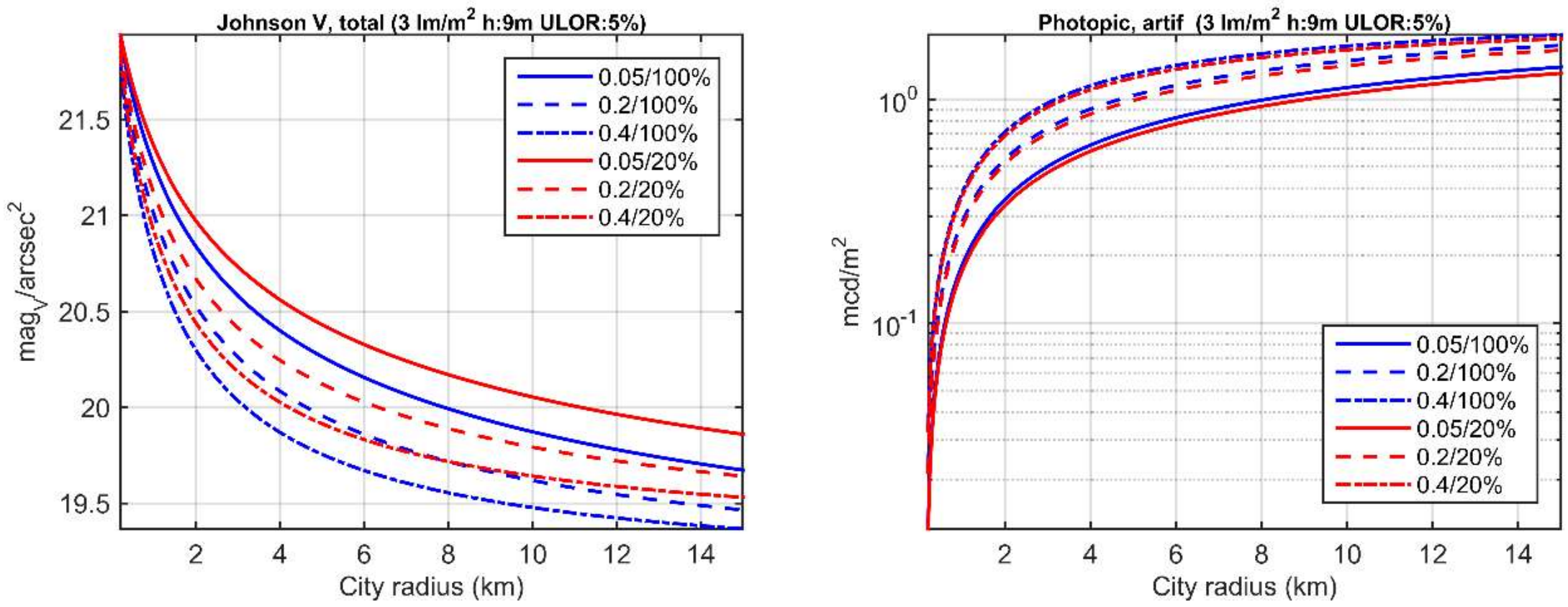

Figure 5: Brightness of the zenith sky for different AOD and LED percent (legend), versus city size. Left: total brightness (including the natural component) in the Johnson V band, in magV/arcsec$^2$. Right: artificial component luminance in mcd/m$^2$. Blue lines: 100% of the light flux in lm comes from 3000 K LED sources. Red lines: mix of 80% high-pressure sodium lamps and 20% LED. Different aerosol concentrations are displayed as different line types (full, dashed, and dash-dotted). See text for details.

Finally, Figure 5 shows the differences between a city with a light inventory composed 100% of 3000 K CCT LED (blue lines) and one that combines an 80% of high-pressure sodium sources (HPS) and a 20% of LED (red lines), where the percents refer to the total emitted light flux in lm, and with ULOR 5%. It can be seen that, irrespectively from the considered aerosol optical depths, cities equipped with a mix of HPS and LED sources are expected to have darker skies than the ones equipped only with 3000 K LEDs. Reduced sky brightnesses can also be achieved by using LED of lower CCT, since it is the blue content of the sources what makes them more harmful for the night sky brightness at short distances, everything else being equal, as is the case analyzed in this paper.

## 4 Discussion

Equation (A12) in the Appendix allows to perform a fast, order-of-magnitude estimation of the night sky brightness attainable in urban settings (or, generally, in any territory with uniform density of light emissions) as a function of the spatial density of the emitted light flux. The results shown in Section 3 strongly suggest that better night skies could be achieved in many cities by means of a judicious choice of the effective areas to be illuminated and of the average illuminance to be used in them. Note that the lumen emission budget considered in this paper, $\Phi_L$, must include the contribution of all existing light sources, including those whose function is not strictly related to roadway illumination, such as scenic lighting, private lighting, LED screens and other. There is no theoretical or procedural difficulty in performing tailored calculations for any particular urban environment without resorting to the simplifications made in this paper: the present calculations just try to answer in general terms the question raised in the title.

Given the unavoidable variability of the atmospheric conditions, any metric for assessing the quality of the night sky has to be conceived and understood in a statistical way. Night sky brightness calculations can be made for a predetermined distribution of atmospheric states and then averaged with appropriate weights or, alternatively, a standard atmospheric state could be chosen by the light pollution community as the reference for calculating the expected sky darkness achievable by keeping the city emissions within acceptable levels. An open debate of this issue would be helpful for reaching an agreement that would enable the comparison of results for different lighting approaches and urban territorial structures. Whether or not a single state of the atmosphere can be adopted as a reasonable worldwide standard or different one should be considered depending on climate areas is still to be determined.

## 5 Conclusions

The darkness of our urban skies can be improved by a sensible choice of the area to be illuminated and the average spatial density of the urban light emissions. The simplified equations derived in this paper can be used for getting quantitative insights on the trade-offs between emissions and starry skies. The results strongly suggest that the complete loss of the starry nights is not an unavoidable fate, even in our large metropolitan areas.

How our city nights should look like is a social and political decision. There is indeed no 'natural' nor 'technically predetermined' way of lighting our nights. The use of artificial light sources creates by definition a new reality different from the natural night, an artificial nightscape whose main features should be collectively decided before choosing the technical solutions that should allow us to fulfil these goals.

It seems feasible, and no doubt desirable, to regain the social control of the nightscape we wish for our cities, with the additional benefit that such measures bring to neighboring regions. Quantitative models like the one described in this paper, and many others more specific and detailed already available in the scientific literature, can be instrumental tools for science-informed public decision-making on the important issue of the preservation of the night in our planet.

## Appendix

### Modelling the brightness of the sky in terms of the city light emissions

In this paper the term "brightness", $B$, is meant to denote the radiance (expressed either in $\mathrm{W \cdot m^{-2} \cdot sr^{-1}}$ or in $\mathrm{photon \cdot s^{-1} \cdot m^{-2} \cdot sr^{-1}}$, depending on the choice of the radiometric system) contained within the spectral sensitivity band $S(\lambda)$ of the detector. $B$ is calculated by integrating the sky spectral radiance $L(\lambda)$, given in units $\mathrm{W \cdot m^{-2} \cdot sr^{-1} \cdot nm^{-1}}$ or $\mathrm{photon \cdot s^{-1} \cdot m^{-2} \cdot sr^{-1} \cdot nm^{-1}}$, within the $S(\lambda)$ band, as

$$B = \int_{0}^{\infty} S(\lambda)L(\lambda)\mathrm{d}\lambda \qquad \text{(A1)}$$

A similar definition can be applied to any sky quality indicator linearly dependent on the spectral radiance at the observer location, including, but not limited to, the zenith radiance, the average hemispheric radiance, the horizontal illuminance, or the average radiance within some angular strip above the horizon [46-48]. The formulation developed below applies equally to any of these indicators, and hence $B$ may be understood as representing any of them; for the particular purpose of this paper, however, we will identify $B$ with the zenith sky brightness.

The spectral bands $S(\lambda)$ analyzed here are the Johnson-Cousins V [26] and the CIE visual photopic sensitivity band $V(\lambda)$ [40]. Their associated photometric systems are defined, by historical reasons, in terms of energies instead of photon numbers, even if the actual interactions of the radiation with the detectors take place on a photon basis. When working with human visual bands in the photopic, scotopic or mesopic adaptation ranges the in-band radiance, expressed in energy units, can be converted into SI luminous units of cd·m$^{-2}$ by multiplying it by the corresponding luminous efficacy coefficients, namely $K_m = 683$ lm·W$^{-1}$ for photopically adapted eyes, $K'_m = 1700$ lm·W$^{-1}$ for scotopically adapted ones, and intermediate values for mesopic adaptation [43].

The spectral radiance $L(\mathbf{r}_0, \lambda)$ at the observer position $\mathbf{r}_0$ can be expressed as a weighted linear combination of the spectral radiant flux emitted by the surrounding light sources. If $\Phi(\mathbf{r}, \lambda)\mathrm{d}^2\mathbf{r}$ is the spectral radiant flux (W·nm$^{-1}$) emitted by the sources contained within the elementary area $\mathrm{d}^2\mathbf{r}$ (m$^2$), such that $\Phi(\mathbf{r}, \lambda)$ is the spectral radiant flux emitted per unit of territory surface (W·m$^{-2}$·nm$^{-1}$), the linear nature of this problem allows to write [63-64]:

$$L(\mathbf{r}_0, \lambda) = \int_{A} \Phi(\mathbf{r}, \lambda)\Psi(\mathbf{r}, \mathbf{r}_0, \lambda)\mathrm{d}^2\mathbf{r} \qquad \text{(A2)}$$

where $\Psi(\mathbf{r}, \mathbf{r}_0, \lambda)$ is the point-spread function (m–2·sr–1) that accounts for the atmospheric propagation and for the angular emission pattern of the sources, assumed to be spectrally and angularly factorable [63], and for the effects related to the terrain (ground and façade reflections, obstacle blocking, etc). The integral is extended to $A$, the relevant territory containing the sources that contribute to the radiance at the observer location. This formal model allows us to use directly the PSF functions proposed by Simoneau et al [58].

As a side note, recall that alternative expressions similar to Eq. (A2) could be written for computing $L(\mathbf{r}_0, \lambda)$ in terms of other inputs. For instance, $\Phi(\mathbf{r}, \lambda)\mathrm{d}^2\mathbf{r}$ could be taken to mean the spectral radiant flux leaving a given patch $\mathrm{d}^2\mathbf{r}$ of the territory, not the spectral flux leaving the luminaires, in which case the ground and façade reflections and the obstacle blocking effects would be included in the values of $\Phi$ rather than in $\Psi$ (that would then basically describe the atmospheric propagation effects). The input $\Phi(\mathbf{r}, \lambda)$ to Eq. (A2) could also be another practical radiometric quantity, for instance the spectral radiance $L(\mathbf{r}, \lambda)$ of the terrain at $\mathrm{d}^2\mathbf{r}$ (W·m$^{-2}$·sr$^{-1}$·nm$^{-1}$), as described in [63], in which case calibrated radiance satellite images could be used as inputs and the PSF will have units m$^{-2}$.

In what follows and for the purposes of the examples presented in this paper we assume that the point-spread function (PSF) $\Psi$ is shift-invariant, that is, spatially homogenous, such that its value depends on the relative position of the source with respect to the observer but not on the absolute positions of them

$$\Psi(\mathbf{r}, \mathbf{r}_0, \lambda) = \Psi(\mathbf{r} - \mathbf{r}_0, \lambda) \qquad \text{(A3)}$$

and we further impose rotational symmetry such that

$$\Psi(\mathbf{r}, \mathbf{r}_0, \lambda) = \Psi(r, \lambda) \qquad \text{(A4)}$$

where $r = \|\mathbf{r} - \mathbf{r}_0\|$ is the source-to-observer distance .

For the purposes of this paper it will be also assumed that the observer is located in an urban pa rk or star observing spot reasonably free from nearby and direct light emission sources within a small zone of radius $R_0$, but surrounded by a wide city or metropolitan area of radius $R$. For a preliminary order-of-magnitude assessment we will further assume that the urban emissions have a uniform spatial density, i.e. that they can be considered homogeneous across the territory when aggregated within each surface integral element $\mathrm{d}^2\mathbf{r}$, so that the spectral flux density does not depend on the position within t he city, and hence $\Phi(\mathbf{r},\lambda)=\Phi(\lambda)$. Under these conditions the radiance at the observer location can be expressed as

$$L(\mathbf{r}_0,\lambda)\equiv L(\lambda;R)=\Phi(\lambda)\int_0^{2\pi}\int_{R_0}^{R}\Psi(r,\lambda)\,r\,\mathrm{d}r\,\mathrm{d}\varphi=2\pi\,\Phi(\lambda)\int_{R_0}^{R}\Psi(r,\lambda)\,r\,\mathrm{d}r \qquad \text{(A5)}$$

Once $L(\lambda;R)$ is known, the zenith sky brightness in the photometric observation band $S(\lambda)$, i.e., t he in-band integrated radiance $B(R)$, dependent on the size of the surrounding light-emitting territory, i s:

$$B(R)=\int_0^{\infty}S(\lambda)L(\lambda;R)\mathrm{d}\lambda=2\pi\int_{R_0}^{R}\int_0^{\infty}S(\lambda)\Phi(\lambda)\Psi(r,\lambda)\,r\,\mathrm{d}r\,\mathrm{d}\lambda \qquad \text{(A6)}$$

and the corresponding value of the anthropogenic brightness of the sky in magnitudes per square arcse cond, $m_{artif}$, is

$$m_{artif}=-2.5\log_{10}\frac{B(R)}{B_{ref}} \qquad \text{(A7)}$$

where $B_{ref}$ is the zero-point of the magnitude scale, that can be arbitrarily chosen but shall be explicit ly specified, associated via Eq. (A1) to a reference spectral radiance $L_{ref}(\lambda)$. Two usual choices to def ine the zero-point of the magnitude scales in astronomy are the spectral irradiance of the star Vega (a Lyr) and the absolute AB irradiance proposed by Oke [65], after expressing them as the radiances of a patch of the sky of size 1 arcsec$^2$ that would produce these irradiances at the entrance of the observi ng instrument [46].

To calculate the observed sky magnitude the contribution of the natural sky brightness shall be ad ded to the numerator of the log argument in Eq. (A7). If the natural sky brightness in the chosen pho tometric band is assumed to be $m_N$, expressed in mag/arcsec$^2$, then the in-band radiance of the natural sky is $B_N=B_{ref}\,10^{-0.4\times m_N}$, and the total brightness in magnitudes per square arcsecond measured by t he observer, $m$, will be

$$m=-2.5\log_{10}\frac{B_N+B(R)}{B_{ref}}=-2.5\log_{10}\left[10^{-0.4\times m_N}+\frac{B(R)}{B_{ref}}\right] \qquad \text{(A8)}$$

Finally, from the above equations one gets:

$$m=-2.5\log_{10}\left[10^{-0.4\times m_N}+\frac{2\pi}{B_{ref}}\int_{R_0}^{R}\int_0^{\infty}S(\lambda)\Phi(\lambda)\Psi(r,\lambda)\,r\,\mathrm{d}r\,\mathrm{d}\lambda\right] \qquad \text{(A9)}$$

In summary, Eq. (A9) provides $m$, the total brightness of the sky in mag/arcsec2 in the $S(\lambda)$ band, ex pressed in a magnitude scale with a zero-point defined by the radiance $B_{ref}$, measured by an observer in the center of an unlit park of radius $R_0$ surrounded by a territory of radius $R$ whose luminaires emi t a spatially averaged homogeneous spectral flux $\Phi(\lambda)$, with the atmospheric conditions, terrain features, and luminaire angular emission patterns accounted for in the shift-invariant and rotationally symmetric PSF $\Psi(r,\lambda)$, when the magnitude of the natural sky is $m_N$.

The spectral radiant flux emitted by the city by unit of urban surface, $\Phi(\lambda)$, can be rewritten in te rms of the spatial density of emitted light, $\Phi_L$, in lm·m$^{-2}$ (equivalent to Mlm·km$^{-2}$) by recalling that t he photopic luminous flux $\Phi_L$ associated to a spectral radiant flux $\Phi(\lambda)$ is:

$$\Phi_L = K_m \int_0^{\infty} V(\lambda)\Phi(\lambda)\mathrm{d}\lambda \quad \text{(A10)}$$

where $V(\lambda)$ is the CIE photopic spectral sensitivity function and $K_m = 683$ lm·W$^{-1}$. Denoting by $\hat{\Phi}(\lambda)$ the normalized radiant flux producing exactly one lumen, we have $\Phi(\lambda) = \Phi_L\,\hat{\Phi}(\lambda)$, where

$$\hat{\Phi}(\lambda) = \frac{\Phi(\lambda)}{K_m \int_0^{\infty} V(\lambda)\Phi(\lambda)\mathrm{d}\lambda} \quad \text{(A11)}$$

has units W·lm$^{-1}$·nm$^{-1}$. Note that the W·lm$^{-1}$ units stem from the $K_m$ factor in the denominator, so for the purpose of calculating $\hat{\Phi}(\lambda)$ it is possible to use a $\Phi(\lambda)$ dataset in arbitrary linear units per nm (they cancel out in Eq. (A11)), which is a convenient option when the shapes of the lamp spectra are known but not their calibrated radiometric flux. With this notation Eq. (A9) becomes

$$m = -2.5\log_{10}\left[10^{-0.4\times m_N} + \frac{2\pi\Phi_L}{B_{ref}}\int_{R_0}^{R}\int_0^{\infty} S(\lambda)\hat{\Phi}(\lambda)\Psi(r,\lambda)\, r\, \mathrm{d}r\, \mathrm{d}\lambda\right] \quad \text{(A12)}$$

Finally, if the fraction of radiant flux emitted towards the lower hemisphere is $\eta = 1 - \mathrm{ULOR}$, and the fraction of the territory surface actually illuminated is $\epsilon$ (encompassing all illuminated city locations, i.e. excluding roofs and other zones where no direct light actually arrives), the average ground illuminance $E$, in lx, is given by $E = \eta\Phi_L/\epsilon$.

## Acknowledgments

This work was supported in part by Xunta de Galicia, grant ED431B 2020/29.